\documentclass{article}
\usepackage{frascatiphys,here,graphicx,subfigure}
\begin{document}
\title{MASSES AND MAGNETIC MOMENTS OF CHARMED BARYONS USING HYPER
CENTRAL MODEL}
\author{
Bhavin Patel, \\
$^*${\em Department of Physics, Sardar Patel University,
  }\\{\em Vallabh Vidyanagar-388 120,Gujarat, India.}\\
Ajay Kumar Rai,         \\
{\em Department of Applied Sciences and Humanities,}\\
{\em SVNIT, Surat-395 007,Gujarat, India.}\\
$^*$P. C. Vinodkumar. \\
 \\
}

\maketitle \baselineskip=11.6pt
\begin{abstract}
Heavy flavour baryons containing one or two charm quarks with
light flavour combinations are studied using the hyper central
description of the three-body system. The confinement potential is
assumed as hyper central coulomb plus power potential with power
index $\nu$. The ground state masses and the magnetic moments of
charmed, $J^P=\frac{1}{2}^+$ and $\frac{3}{2}^+$ baryons are
computed for different power index, $ \nu$  starting from 0.5 to
2.0.
 \end{abstract}
\baselineskip=14pt
\section{Introduction}
The investigation of properties of hadrons containing heavy quarks
is of great interest in understanding the dynamics of QCD at the
hadronic scale. There is renewed interest both experimentally and
theoretically in the static properties of heavy flavour baryons
such as  mass and magnetic moments \cite{Giannini2001,
 Ebert2002,Amand2006,Athar2005,Bha_arx2007}.
  Many of the narrow hadron resonances observed recently, brought up
 surprises in QCD spectroscopy\cite{Rosner609}.
Recent years, experimental facilities at Belle, BABAR, DELPHI,
CLEO, CDF etc have been successful in discovering heavy baryon
states along  with other heavy flavour mesonic states
\cite{Athar2005,Mizuk2005,Aubert2007,Gorelov2007} and  many new
states are expected in near future. Baryons are not only the
interesting systems to study the quark dynamics and their
properties but are also interesting from the point of view of
simple systems to study three body problems.  All these reasons
make the study of heavy flavour spectroscopy extremely important
and interesting. Here, we employ the hyper central approach to
study the three-body problem, particularly the baryons
constituting one or two charm quarks. In the present study, the
magnetic moments of heavy flavour baryons are computed based on
 nonrelativistic quark model.
\section{Hyper Central scheme for baryons} 
\begin{figure}[H]
    \begin{center}
        {\includegraphics[scale=0.56]{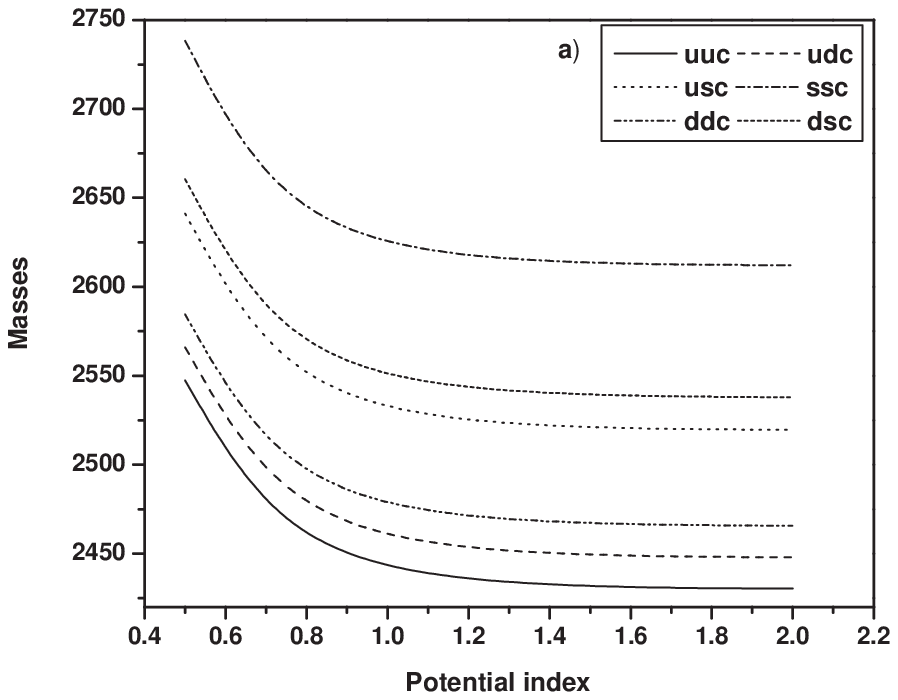}}
        {\includegraphics[scale=0.56]{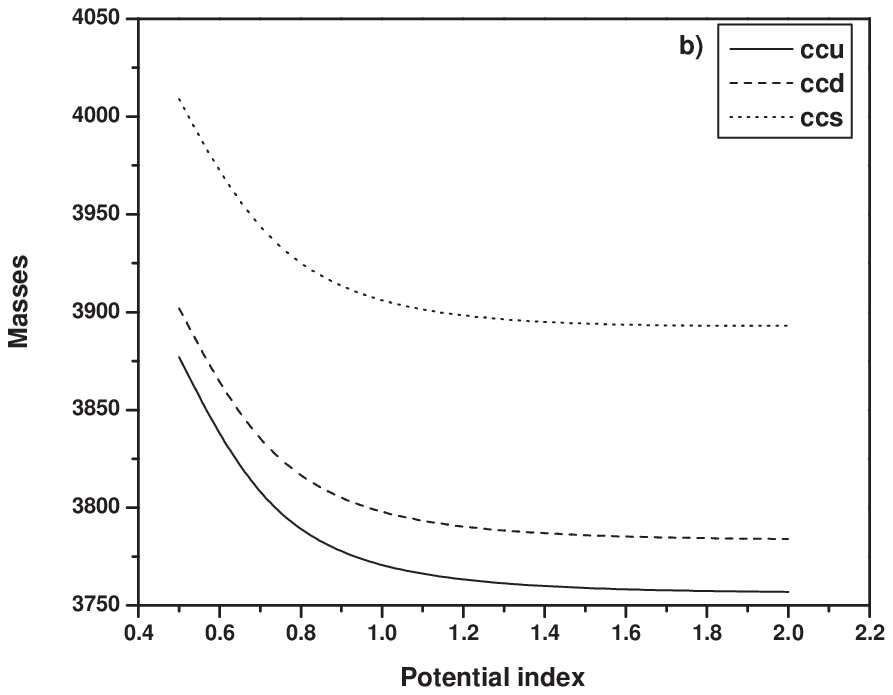}}
        \caption{\it Variation of spin
average masses with potential index $\upsilon$ for single charm
baryons [a] and double charm baryons [b].} \label{fig:massvspot}
    \end{center}
\end{figure}

Quark model
description of baryons is a simple three body system of interest.
The Jacobi co-ordinates to describe baryon as a bound state of
three constituent quarks are given by \cite{Simonov1966}
\begin{equation}\label{}
   \vec{\rho}=\frac{1}{\sqrt{2}}(\vec{r}_1-\vec{r}_2)\,\,{;}\,\,
 \vec{\lambda}=\frac{1}{\sqrt{6}}(\vec{r}_1+\vec{r}_2-2\vec{r}_3)
\end{equation}
Further, defining the hyper spherical coordinates which are given
by the angles $\Omega_\rho=(\theta_\rho,\phi_\rho)\,\,{;}\,\,$
$\Omega_\lambda=(\theta_\lambda,\phi_\lambda)$ together with the
hyper radius, $x$ and hyper angle $\xi$ respectively as,
$x=\sqrt{\rho^2+\lambda^2}\,\,{;}\,\,\xi=\arctan\left(\frac{\rho}{\lambda}\right)$,
the model Hamiltonian for baryons can be written as
\begin{equation}\label{eq:404}
H=\frac{P^2_\rho}{2\,m_\rho}+\frac{P^2_\lambda}{2\,m_\lambda}+V(\rho,\lambda)=\frac{P^2_x}{2\,m}+V(x)\\
\end{equation}\label{}
Here the potential $V(x)$ is not purely a two body interaction but
it contains three-body effects also. The reduced mass $m$ is
defined as $m=\frac{2\,\,m_\rho\,\, m_\lambda}{m_\rho+m_\lambda}$,
where $m_\rho=\frac{2\,\,m_1\,\, m_2}{m_1+m_2}
 \,\,{;}\,\,
m_\lambda=\frac{3\,\,m_3\,\,(m_1+m_2)}{2\,(m_1+m_2+m_3)}$ and
$m_1$, $m_2$ and $m_3$ are masses of the three constituent quarks.
For the present study,  we consider the hyper central potential
$V(x)$ as\, \cite{Bha_arx2007}
\begin{equation}\label{eq:410}
V(x)=-\frac{\tau}{x}+\beta x^\upsilon+\kappa+A\,\,e^{-\alpha
x}\sum\limits_{i{\neq}j}\sigma_i \cdot
\sigma_j\\
\end{equation}\label{}
In this hyperspherical representation of the potential we consider
$\tau=\frac{2}{3}\,b\,\alpha_s$, $\beta\approx m\tau$,
$\kappa=(\sqrt{2}-1)\alpha_{s}N_{c}N_{f} m $, where $N_{c}=3.0$
while $N_{f}=4$ for charmed baryons and other model parameters are
given in\cite{Bha_arx2007}. The trial wave function is taken
\cite{Giannini2001}as the hyper coulomb radial wave function. We
study the baryons having different choice of the light
flavour(q)combinations for qqc and ccq systems. The computed spin
average mass with respect to the potential index, $\nu$ are shown
in Fig \ref{fig:massvspot}. The computed masses of the spin
$\frac{1}{2}$ and $\frac{3}{2}$, single charm and double charm
baryons are given in Tables \ref{tab:01} and \ref{tab:02}
respectively.

\begin{table}
\begin{center} \caption{\it Single charm baryon masses(masses are in
MeV)} \vspace{0.001in} \label{tab:01}
\begin{tabular}{|l|l|l|l|l|l|}
\hline
 Baryon &P.I.($\nu$)
&{$\textbf{J}^P=\frac{1}{2}^+$}&Others&{$\textbf{J}^P=\frac{3}{2}^+$}&Others\\
\hline \hline
$\Sigma^{++}_{c}$&0.5&2550&2453\cite{Amand2006}&2618&$-$\\
(uuc)&0.7&2473&2454$\pm$0.18\cite{PDG2006}&2538&2518$\pm$0.6\cite{PDG2006} \\
&1.0&2443&2460$\pm$80\cite{Bowler1998}&2506&2440$\pm$70\cite{Bowler1998} \\
&1.5&2436&&2499& \\
&2.0&2436&&2498& \\
 $\Sigma^{+}_{c} $&0.5&2568&2451\cite{Amand2006}& 2638&$-$\\
(udc)&0.7&2491&2439\cite{Ebert2002}& 2557&2518\cite{Ebert2002}\\
&1.0&2460&2453\cite{Roncaglia1995}&2525&2520\cite{Roncaglia1995}\\
&1.5&2454&2452\cite{Mathur2002}&2518&2538\cite{Mathur2002}\\
&2.0&2453&2448\cite{Garcilazo2007}&2517&2505\cite{Garcilazo2007}\\
&&&$2453\pm0.4$\cite{PDG2006}&&$2518\pm2.3$\cite{PDG2006}\\

 $\Sigma^{0}_{c} $&0.5&2586&2452
\cite{Amand2006}&2658
&$-$\\
(ddc)&0.7&2508&2454$\pm$0.18\cite{PDG2006}& 2577&2518$\pm$0.5\cite{PDG2006}\\
&1.0& 2477&& 2544&\\
&1.5& 2471&&2537 &\\
&2.0& 2470&& 2537&\\
 $\Xi^{+}_{c} $&0.5&2642&2466\cite{Amand2006}& 2720&$-$\\
(usc)&0.7&2561&2481\cite{Ebert2002}& 2636&2654\cite{Ebert2002}\\
&1.0&2530&2468\cite{Roncaglia1995}& 2603&2650\cite{Roncaglia1995}\\
&1.5&2523&2473\cite{Mathur2002}&2596&2680\cite{Mathur2002}\\
&2.0&2523&2496\cite{Garcilazo2007}&2595&2633\cite{Garcilazo2007}\\
&&&2468$\pm$0.4\cite{PDG2006}&&2647$\pm$1.4\cite{PDG2006}\\
&&&2410$\pm$50\cite{Bowler1998}&&2550$\pm$80\cite{Bowler1998}\\
$\Xi^{0}_{c}$ &0.5&2653&$2472$\cite{Amand2006} &2734&$-$\\
(dsc)&0.7&2579 &2471$\pm$0.4\cite{PDG2006}&2656&2646$\pm$1.2\cite{PDG2006}\\
&1.0&2548&&2623&\\
&1.5&2541&&2616&\\
&2.0&2541&&2615&\\
$\Omega^{0}_{c}
$&0.5&2720&2698\cite{Amand2006}& 2810&$-$\\
(ssc)&0.7&2652&2698\cite{Ebert2002}& 2739&2768\cite{Ebert2002}\\
&1.0& 2620&2710\cite{Roncaglia1995}&2704&2770\cite{Roncaglia1995}\\
&1.5&2613&2678\cite{Mathur2002}&2697&2752\cite{Mathur2002}\\
&2.0&2613&2701\cite{Garcilazo2007}&2697&2759\cite{Garcilazo2007}\\
&&&2680$\pm$70\cite{Bowler1998}&&2660$\pm$80\cite{Bowler1998}\\
&&&2698$\pm$2.6\cite{PDG2006}&&\\
\hline
\end{tabular}
\end{center}
\end{table}

\begin{table}[t]
\begin{center} \caption{\it Doubly charm baryon masses (masses are in MeV)}
\vskip 0.01 in \label{tab:02}
\begin{tabular}{|l|l|l|l|l|l|}
\hline  Baryon & P.I.($\nu$)
&{$\textbf{J}^P=\frac{1}{2}^+$}&Others&{$\textbf{J}^P=\frac{3}{2}^+$}&Others\\
\hline \hline $\Xi^{++}_{cc}$&0.5&3838&$3612^{+17}$\cite{Albertus2006}& 3915&$3706^{+23}$\cite{Albertus2006}\\
(ccu)&0.7&3760&3620\cite{Ebert2002}& 3833&3727\cite{Ebert2002}\\
&1.0&3730&3480\cite{Kselev2002}&3800&3610\cite{Kselev2002}\\
&1.5&3723&3740\cite{Tong2000}&3792&3860\cite{Tong2000}\\
&2.0&3723&3478\cite{Gershtein2002}&3792&3610\cite{Gershtein2002}\\
 $\Xi^{+}_{cc}$&0.5&3862&3605$\pm$23\cite{Lewis2001}&3945&3685$\pm$23\cite{Lewis2001}\\
(ccd)&0.7&3786&3620\cite{Ebert2002}&3862 &3727\cite{Ebert2002}\\
&1.0&3755&3480\cite{Kselev2002}&3828&3610\cite{Kselev2002}\\
&1.5&3748&3740\cite{Tong2000}&3821&3860\cite{Tong2000}\\
&2.0&3748&3478\cite{Gershtein2002}&3820&3610\cite{Gershtein2002}\\
 $\Omega^{+}_{cc} $&0.5&3962&$3702^{+41}$\cite{Albertus2006}& 4056&$3783^{+22}$\cite{Albertus2006}\\
(ccs)&0.7&3889 &3778\cite{Ebert2002}& 3978&3872\cite{Ebert2002}\\
&1.0&3857&3590\cite{Kselev2002}&3944&3690\cite{Kselev2002}\\
&1.5&3850&3760\cite{Tong2000}&3936&3900\cite{Tong2000}\\
&2.0&3850&3590\cite{Gershtein2002}&3936&3690\cite{Gershtein2002}\\
&&&3733$\pm09$\cite{Lewis2001}&&3801$\pm09$\cite{Lewis2001}\\
\hline
\end{tabular}\vskip 0.001 in
\end{center}
\end{table}

\section{Effective quark mass and Magnetic moments of heavy baryons}
We define an effective mass for the bound quarks within the baryon
as
\begin{equation}\label{eq:417} m^{eff}_{i}=m_i\left(
1+\frac{<H>}{\sum\limits_{i}m_{i}}\right) \\
\end{equation}
and we express the magnetic moment of baryons in terms of its
constituent quarks as
\begin{equation}
\mu_B=\sum\limits_{i}\left<\phi_{sf}\mid\frac{e_{i}}{2m_{i}^{eff}}
\vec{\sigma}_{i}\mid\phi_{sf}\right>
\end{equation}
Here $e_{i}$ and $\sigma_{i}$ represents the charge and the spin
of the quark constituting the baryonic state and $\mid\phi_{sf}>$
represents the spin-flavour wave function of the respective
baryonic state. Extending the SU(2)$_S$ $\times$ $SU(3)_f$ spin
flavour structure of the light flavour sector\cite{Simonov2002}in
SU(2)$_S$ $\times$ $SU(4)_f$ spin flavour structure with charm, we
compute the magnetic moments of the spin$\frac{1}{2}$ and
spin$\frac{3}{2}$  charmed baryons.  Our results are listed in
Tables \ref{tab:03} and \ref{tab:04} respectively.
\begin{table}[t]
 \caption{\it Magnetic moments of single charm baryons
 in terms of  $\mu_{N}$}\vskip 0.1 in
 \label{tab:03}
\begin{tabular}{|l|c|c|c|c|c|c|c|} \hline
&\multicolumn{5}{c}{\textbf{\underline{\, \,\,\,\, \,\,\,\, \,\,\,\, \,\,\,\, \,\,\,Potential index $\nu$\, \, \,\,\,\, \,\, \, \,}}}&&\\
Baryon &0.5&0.7&1.0&1.5&2.0&RQM&NRQM\\
\hline \hline
$\Sigma^{++}_{c}$&1.8809&1.9394&1.9635&1.9688&1.9692&1.76&1.86\\
$\Sigma^{*++}_{c}$   &3.2806&3.3837&3.4272&3.4373&3.4379&$-$&$-$\\
$\Sigma^{+}_{c}$&0.3959&0.4082&0.4133&0.4144&0.4144&0.36&0.37\\
$\Sigma^{*+}_{c}$   &1.1092&1.1441&1.1588&1.1622&1.1624&$-$&$-$\\
$\Xi^{+}_{c}$&0.4542&0.4684&0.4742&0.4755&0.4755&0.41&0.37\\
$\Xi^{*+}_{c}$   &1.1893&1.2269&1.2425&1.2460&1.2463&$-$&$-$\\
$\Omega^{0}_{c}$&-0.9612&-0.9860&-0.9981&-1.0006&-1.0008&-0.85&-0.85\\
$\Omega^{*0}_{c}$   &-0.8703&-0.8931&-0.9044&-0.9068&-0.9070&$-$&$-$\\
$\Sigma^{0}_{c}$&-1.0854&-1.1193&-1.1332&-1.1362&-1.1364&-1.04&-1.11\\
$\Sigma^{*0}_{c}$   &-1.0540&-1.0873&-1.1012&-1.1044&-1.1046&$-$&$-$\\
$\Xi^{0}_{c} $&-1.0231&-1.0526&-1.0655&-1.0683&-1.0685&-0.95&-0.98\\
$\Xi^{*0}_{c} $   &-0.9618&-0.9898&-1.0024&-1.0052&-1.0054&$-$&$-$\\
\hline
\end{tabular}
RQM\cite{Amand2006}, NRQM\cite{Amand2006}
\end{table}
\begin{table}[t]
 \caption{\it Magnetic moments of doubly charm baryons in terms of  $\mu_{N}$}\vskip 0.1 in
 \label{tab:04}
\begin{tabular}{|l|c|c|c|c|c|c|c|}
\hline
&\multicolumn{5}{l}{\textbf{\underline{\, \,\,\, \,\,\,\, \,\,\,\, \,\,\,\, \,\,\,\, \,\,\,Potential index $\nu$\, \,\,\,\, \,\,\,\, \,\,\,\, \,\,\,\,}}}&&\\
Baryon &0.5&0.7&1.0&1.5&2.0&NRQM &AL1 \\
\hline \hline
 $\Xi^{++}_{cc}$&-0.0151&-0.0154&-0.0156&-0.0156&-0.0156&-0.01&-0.208\\
$\Xi^{*++}_{cc}$   &2.1934&2.2406&2.2602&2.2646&2.2649&$-$&$2.670$\\
 $\Omega^{+}_{cc}$&0.6910&0.7040&0.7098&0.7110&0.7110&$0.67$&0.635\\
$\Omega^{*+}_{cc}$   &0.0908&0.0926&0.0934&0.0936&0.0936&$-$&0.139\\
 $\Xi^{+}_{cc}$&0.7279&0.7426&0.7487&0.7500&0.7501&0.74&0.785\\
$\Xi^{*+}_{cc} $   &0.0031&0.0031&0.0032&0.0032&0.0032&$-$&-0.311\\
 \hline
\end{tabular}
NRQM\cite{Amand2006}, AL1\cite{Albertus2006}
\end{table}

\section{Results and Discussion}
We have employed the hyper central model with hyperspherical
potential of the coulomb plus power potential to study the masses
and magnetic moments of baryons containing one or two charm
flavour quarks. It is important to see that the baryon mass do not
change appreciably for the potential power index $\nu
> 1. 0$ (See Fig \ref{fig:massvspot}). The model parameters are fixed for
this saturated value to the experimental spin average mass of the
$\Sigma_c^*(2518)-\Sigma_c(2454)$(udc) system \cite{PDG2006} and
the spin hyperfine parameter is fixed to yield their mass
difference of 64 MeV \cite{PDG2006}. All other baryonic masses are
predicted without changing any of these parameters. It is
interesting to note that our results are in fair agreement with
existing experimental as well as other theoretical model
predictions.  The result of single charm baryons are in accordance
with the lattice results as well as with other model predictions.
The mass variations of the single charm baryons with respect to
$\nu$ from 0.5 to 2.0 are found to be around 100 $MeV$ only. \\

 Our predictions on the doubly charm
baryons are compared with other theoretical model predictions in
Table \ref{tab:02}. Since there are larger disagreement among the
different model predictions, only the future experiments on these
doubly charm baryons would be able to test the validity of the
theoretical model predictions. However, the hyperfine splitting of
76.6 $MeV$ for $\Xi_{cc}^{*}-\Xi_{cc}$ obtained from Lattice
predictions\cite{Lewis2001} is very close to our calculations of
73 $MeV$. It is important to note that the predictions of the
magnetic moment of all the heavy hadrons studied here are with no
free parameters. Our results for the magnetic moment of single
charm baryons with spin $\frac{1}{2}$ are in accordance with the
predictions of the full treatment of relativistic quark model as
well as with the nonrelativistic approximation reported in
\cite{Amand2006}. In the case of doubly charm baryons,  our
predictions for both $J=\frac{1}{2}$ and $\frac{3}{2}$ baryons are
found to be in good accordance with the recent predictions based
on a potential model, AL1 \cite{Albertus2006} and NRQM
\cite{Amand2006}
results [See Table \ref{tab:04}].\\

We conclude that the three body interaction assumed in our model
plays a significant role in the description of heavy flavour
baryonic properties in particular their masses and magnetic
moments. The behavior of the masses against the potential index
$\nu$ as shown Fig. \ref{fig:massvspot} indicates saturation of
the basic quark-quark interactions within the heavy baryons as the
potential index $\nu > 1.0$. We hope that, the predicted many of
the baryonic ground state will be observed in the future
high luminosity heavy flavour experiments.\\

{\bf Acknowledgement:} Part of this work is done with a
 financial support from University Grant
commission, Government of India under a Major research project
\textbf{F. 32-31/2006 (SR)}.

\end{document}